\documentclass{article}
\usepackage{amsmath, amssymb, graphicx, hyperref}
\usepackage[utf8]{inputenc}
\usepackage[accepted]{icml2025}

\title{FinRL-DeepSeek: LLM-Infused Risk-Sensitive Reinforcement Learning for Trading Agents}

\author{
    Mostapha Benhenda \\
\texttt{mostaphabenhenda@gmail.com}
}

\date{}

\begin{document}

\maketitle

\begin{abstract}
This paper presents a novel risk-sensitive trading agent combining reinforcement learning and large language models (LLMs). We extend the Conditional Value-at-Risk Proximal Policy Optimization (CPPO) algorithm, by adding risk assessment and trading recommendation signals generated by a LLM from financial news. Our approach is backtested on the Nasdaq-100 index benchmark, using financial news data from the FNSPID dataset and the DeepSeek V3, Qwen 2.5 and Llama 3.3 language models. The code, data, and trading agents are available at:  
\url{https://github.com/benstaf/FinRL_DeepSeek}

\end{abstract}

\section{Introduction}
Automated trading agents increasingly leverage reinforcement learning (RL), but often overlook two aspects:
\begin{enumerate}
    \item Integration of alternative data sources, such as financial news.
    \item Risk management.
\end{enumerate}

This paper introduces a novel hybrid RL-LLM trading agent, incorporating financial news insights at both the action and risk levels. The main contribution of the paper is the introduction of an LLM-based risk assessment score derived from news, in addition to LLM-based trading recommendations. This demonstrates the potential of LLMs for feature extraction from news beyond standard sentiment analysis, achieved through carefully designed prompts.

\section{Related Work}

There is a growing body of work on LLM-augmented RL agents, such as \cite{ananya2,openreview25}. Compared with them, the way LLMs influence the RL agent is simpler here.

In FinGPT \cite{fingpt}, RL is applied to the training of LLM on stock prices. This approach is more complex than ours, which has the benefit of leveraging established LLM APIs.

Other hybrid RL-LLM approaches include FinCon \cite{fincon}, which makes a complex distillation of information through a synthesized multi-agent collaboration mechanism. In this paper, we only use simple prompts.

These RL-LLM approaches also differ from "pure LLM" agents like \cite{lopezlira2024} and \cite{tradexpert} that only use LLM recommendations.

\section{Data and LLM Prompts}

We use the FNSPID dataset \cite{fnspid}, which contains 15.7 million time-aligned financial news records spanning 1999–2023. To reduce LLM API costs, we randomly select a representative news article per stock per day, reducing the dataset to 2 million records. Using LLMs (DeepSeek V3 \cite{deepseekv3}, Qwen 2.5 72B \cite{qwen25} Llama 3.3 70B \cite{llama3}), we extract stock recommendations and risk assessment using the following prompts :
\bigskip

\textbf{Stock Recommendation Prompt (from \cite{fnspid}):}
\begin{quote}
    "You are a financial expert with stock recommendation experience. Based on a specific stock, score for range from 1 to 5, where 1 is negative, 2 is somewhat negative, 3 is neutral, 4 is somewhat positive, 5 is positive"
\end{quote}

\textbf{Risk Assessment Prompt (new):}
\begin{quote}
    "You are a financial expert specializing in risk assessment for stock recommendations. Based on a specific stock, provide a risk score from 1 to 5, where: 1 indicates very low risk, 2 indicates low risk, 3 indicates moderate risk (default if the news lacks any clear indication of risk), 4 indicates high risk, and 5 indicates very high risk."
\end{quote}

\section{Trading Algorithms}

\subsection{Reinforcement Learning agents on price-only data}

\subsubsection{Proximal Policy Optimization (PPO)}

OpenAI \cite{PPO} introduced Proximal Policy Optimization (PPO) reinforcement learning algorithm,  and \cite{ssrnppo,finrl} applied it to stock trading 

The PPO objective ensures stable policy updates by clipping the probability ratio:
\[
L_{\text{PPO}}(\theta) = \mathbb{E}\left[\min\left(r_t(\theta) \cdot A_t, \text{clip}\left(r_t(\theta), 1 - \epsilon, 1 + \epsilon\right) \cdot A_t\right)\right]
\]
where:
\begin{itemize}
    \item \( r_t(\theta) = \frac{\pi_\theta(a_t | s_t)}{\pi_{\theta_{\text{old}}}(a_t | s_t)} \) is the probability ratio between the new policy \( \pi_\theta \) and the old policy \( \pi_{\theta_{\text{old}}} \).
    \item \( A_t \) is the advantage estimate at time \( t \).
    \item \( \epsilon \) is the clipping parameter that restricts large updates to the policy.
\end{itemize}

\subsubsection{Conditional Value at Risk-Proximal Policy Optimization (CVaR-PPO)}

Conditional Value at Risk-Proximal Policy Optimization (CVaR-PPO) \cite{cppo} extends PPO with a risk constraint to penalize high-loss trajectories:  
\begin{multline}
L_{\text{CVaR-PPO}}(\theta, \eta, \lambda) = L_{\text{PPO}}(\theta) + \\ \lambda \left( \frac{1}{1-\alpha} \mathbb{E}[(\eta - D(\pi_\theta))^+] - \eta + \beta \right)
\end{multline}

where:
    \begin{itemize}
        \item \( L_{\text{PPO}}(\theta) \) is the PPO objective.
        \item \( D(\pi_\theta) \) is the trajectory return.
        \item \( \eta \) is the CVaR threshold.
        \item \( (\eta - D(\pi_\theta))^+ = \max(0, \eta - D(\pi_\theta)) \) represents the CVaR loss beyond the threshold.
        \item \( \lambda \) is the Lagrange multiplier enforcing the constraint.
        \item \( \alpha \) is the CVaR confidence level (e.g., 0.05 for the worst 5\%).
        \item \( \beta \) is an auxiliary penalty parameter.
    \end{itemize}

To our knowledge, we are the first to apply CVaR-PPO to stock trading.

\subsection{LLM-infused PPO}

In LLM-infused PPO, we compute stock-specific recommendation scores (\(S_f\)) from the FNSPID dataset using an LLM. These scores influence trading actions. The influenced action is:  
\[
a_t^{\text{mod}} = S_f \cdot a_t
\]  
where:

\begin{itemize}
    \item \( S_f > 1 \): Amplifies actions under positive recommendation.
    \item \( S_f < 1 \): Dampens actions under negative recommendation.
    \item \( S_f = 1 \): Leaves actions unchanged.
\end{itemize}

This perturbation \( S_f \) is chosen close to 1 to keep stability of the algorithm. For example:
\begin{itemize}
    \item \( S_f = 1.1 \) if stock score is 5 and \( a_t > 0 \) or score is 1 and \( a_t < 0 \)
    \item \( S_f = 1.05 \) if stock score is 4 and \( a_t > 0 \) or score is 2 and \( a_t < 0 \)
    \item \( S_f = 0.95 \) if stock score is 4 and \( a_t < 0 \) or score is 2 and \( a_t > 0 \)
    \item \( S_f = 0.9 \) if stock score is 5 and \( a_t < 0 \) or score is 1 and \( a_t > 0 \)
\end{itemize}
\subsection{LLM-infused CVaR-PPO(CPPO)}
Financial news from the FNSPID dataset is processed with an LLM to generate risk scores (\(R_f^i\)) for each stock $i$ and day. 

The perturbation \( R_f^i \) is also chosen close to 1 to keep stability of the algorithm. For example:
\begin{itemize}
    \item \( R_f^i = 1.1 \) for very high risk score 5
    \item \( R_f^i = 1.05 \) for high risk score 4
    \item \( R_f^i = 1 \) for moderate risk score 3
    \item \( R_f^i = 0.95 \) for low risk score 2
    \item \( R_f^i = 0.9 \) for very low risk score 1
\end{itemize}

The aggregate risk score \( R_f \) is defined as:
\[ R_f = \sum_i w_i R_f^i \]
where \( w_i \) is the financial weight of stock \( i \) in the portfolio (we have: \( \sum_i w_i = 1 \)).

These scores adjust trajectory returns in Cvar PPO to reflect market risk. The adjusted return is defined as:  
\[
D_{R_f}(\pi_\theta) = R_f \cdot D(\pi_\theta)
\]  

At the end, financial news influence trading actions in two ways: through $S_f$ and $R_f$.

\section{Results}

\subsection{Early Stopping: 400-500k training steps}

In these tests, we use a 10\% infusion of LLM, which means that the original PPO and CPPO up to 10\%

\begin{figure}
    \centering
    \includegraphics[width=1\linewidth]{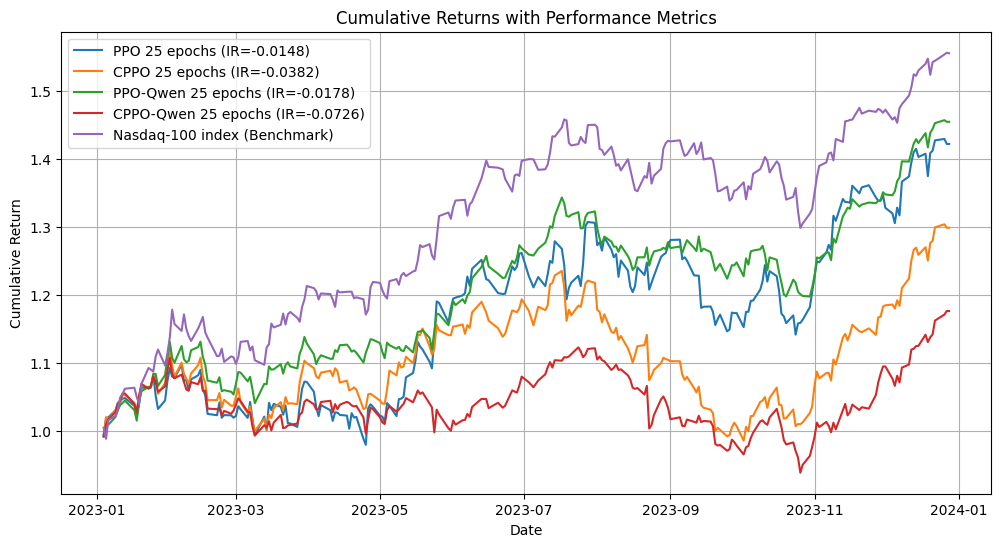}
    \caption{Backtesting after 500k training steps, 3 years training history, 1 year trading}
    \label{fig1}
\end{figure}

\subsubsection{Figure \ref{fig1}:}

\textbf{Training Period:} 2019–2022 \\
\textbf{Training Iterations:} 500k steps (25 epochs, 20k steps each) \\
\textbf{Trading Period:} 2023

These preliminary results in figure 1 with Qwen 2.5 show that integrating LLM stock recommendations consistently improves PPO in terms of cumulative returns.
However, it's not sufficient to beat Nasdaq 100 benchmark so far.

\begin{figure}
    \centering
    \includegraphics[width=1\linewidth]{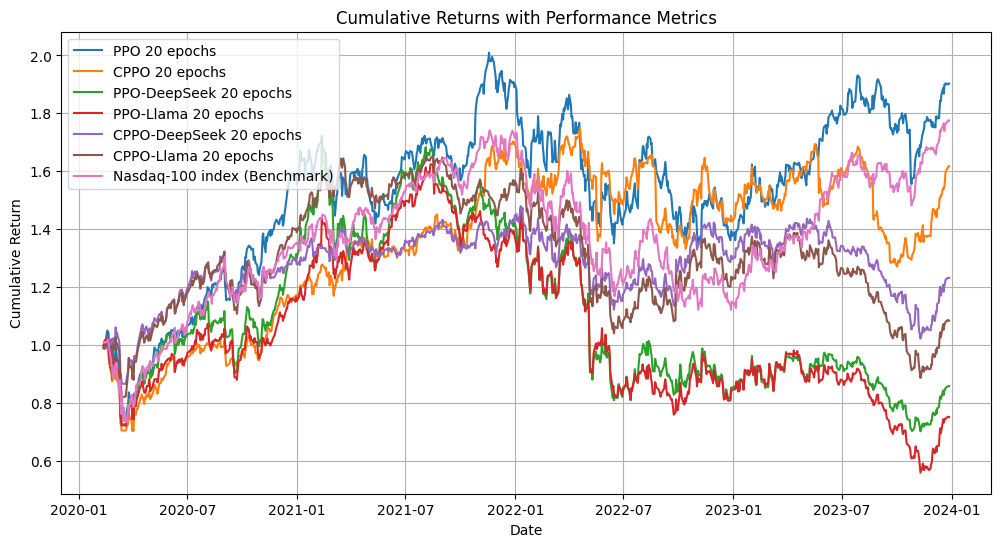}
    \caption{Backtesting after 400k training steps, 6 years training history, 3 years trading}
    \label{fig2}
\end{figure}

\subsubsection{Figure \ref{fig2}:}

\textbf{Training Period:} 2013–2018 \\
\textbf{Training Iterations:} 400k steps (20 epochs, 20k steps each) \\
\textbf{Trading Period:} 2019-2023

Results show significant improvement when taking longer training period (6 years vs. 3 years) for PPO and CPPO
However, PPO remains too volatile. 

DeepSeek V3 is slightly better than Llama 3.3 all other things being equal.

The use of LLM always worsens performance in this test.

\subsection{After training for 2 Million steps}

\begin{figure}
    \centering
    \includegraphics[width=1\linewidth]{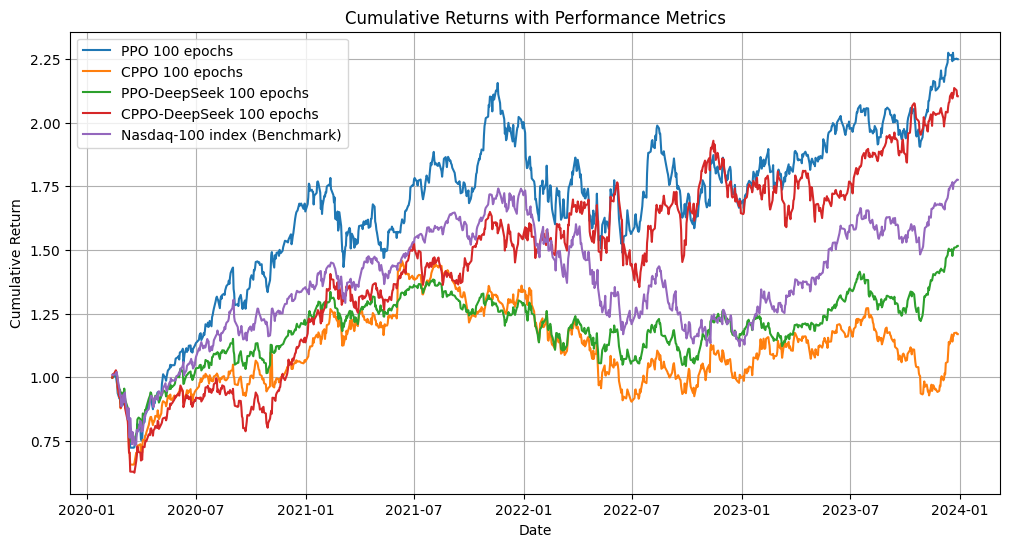}
    \caption{After training for 2 Million steps (100 epochs, 20k steps each)}
    \label{fig5}
\end{figure}

\subsubsection{Figure \ref{fig5}:}

\begin{table}[h]
    \centering
    \resizebox{\columnwidth}{!}{%
    \begin{tabular}{lccc}
        \hline
        Model & Information Ratio & CVaR & Rachev Ratio \\
        \hline
        PPO (100 epochs) & 0.0100 & -0.0394 & 1.0637 \\
        CPPO (100 epochs) & -0.0148 & -0.0439 & 1.0404 \\
        PPO-DeepSeek (100 epochs) & -0.0093 & -0.0338 & 0.9890 \\
        CPPO-DeepSeek (100 epochs) & 0.0078 & -0.0437 & 0.9818 \\
        \hline
    \end{tabular}%
    }
    \caption{Performance metrics for different models over 100 epochs.}
    \label{tab:metrics}
\end{table}

\begin{figure}
    \centering
    \includegraphics[width=1\linewidth]{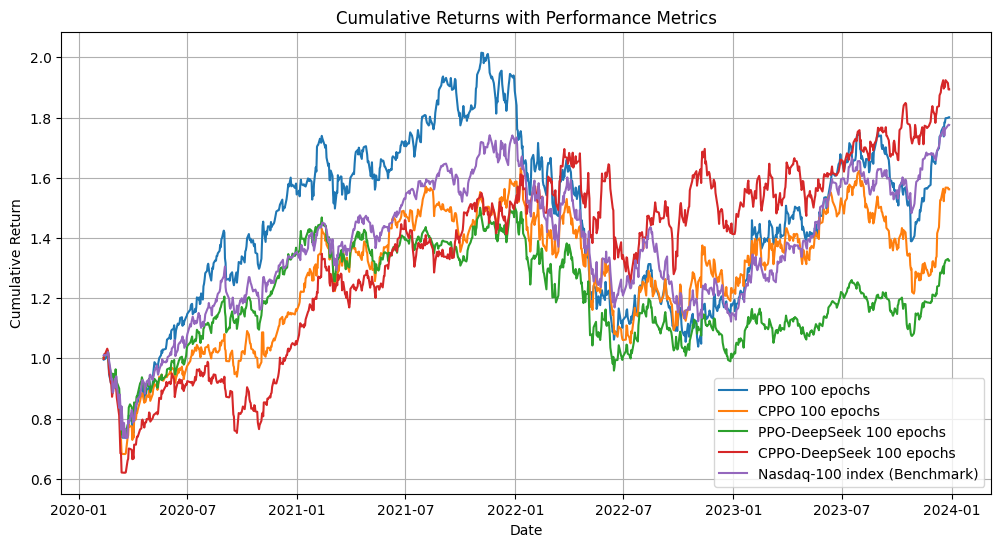}
    \caption{After training for 2 Million steps (100 epochs, 20k steps each) - Second run}
    \label{fig3}
\end{figure}

\subsubsection{Figure \ref{fig3}:}

In these two runs (RL agents are stochastic), PPO and CPPO-DeepSeek outperform other methods, including the Nasdaq 100 benchmark. 

PPO seems to perform better in Bull markets, and CPPO-DeepSeek in Bear markets, the transition being at the end of 2021 before the war in Ukraine and the crisis that followed.

\subsection{Impact of LLM infusion strength}

In the following tests, we tweak LLM infusion strength parameters from 10\% down to 0.1\% (i.e. LLM perturbation parameters vary from $0.9-1.1$ to $0.999-1.001$)

\begin{figure}
    \centering
    \includegraphics[width=1\linewidth]{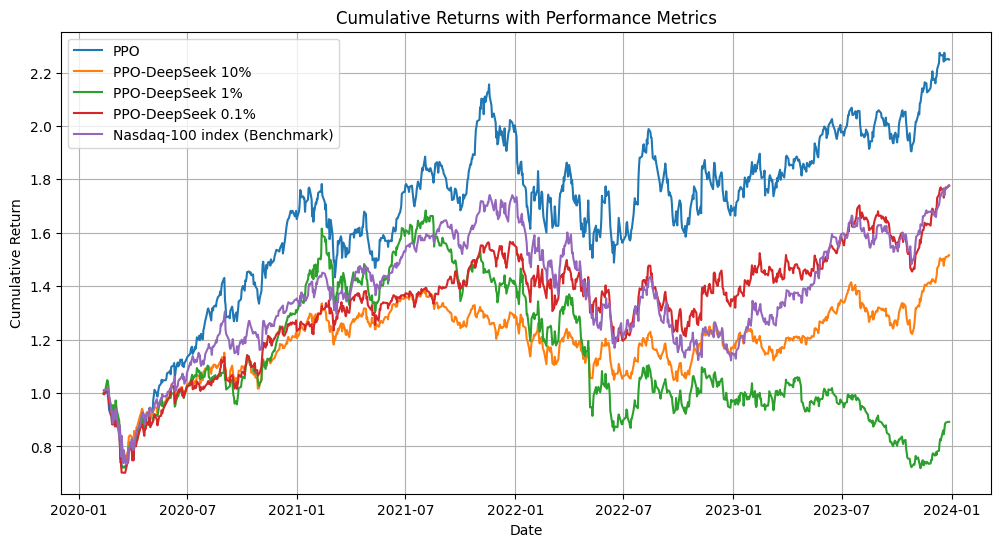}
    \caption{Impact of LLM infusion on PPO}
    \label{fig7}
\end{figure}

\subsubsection{Figure \ref{fig7}:}

\begin{table}[h]
    \centering
    \resizebox{\columnwidth}{!}{%
    \begin{tabular}{lccc}
        \hline
        Model & Information Ratio & CVaR & Rachev Ratio \\
        \hline
        PPO & 0.0100 & -0.0394 & 1.0637 \\
        PPO-DeepSeek 10\% & -0.0093 & -0.0338 & 0.9890 \\
        PPO-DeepSeek 1\% & -0.0252 & -0.0459 & 1.0394 \\
        PPO-DeepSeek 0.1\% & -0.0011 & -0.0375 & 0.9536 \\
        \hline
    \end{tabular}%
    }
    \caption{Performance metrics for PPO and PPO-DeepSeek variants.}
    \label{tab:ppo_metrics}
\end{table}

For PPO-DeepSeek, stronger LLM infusion mostly degrades performance, even for tiny (0.1\%) perturbations.

\begin{figure}
    \centering
    \includegraphics[width=1\linewidth]{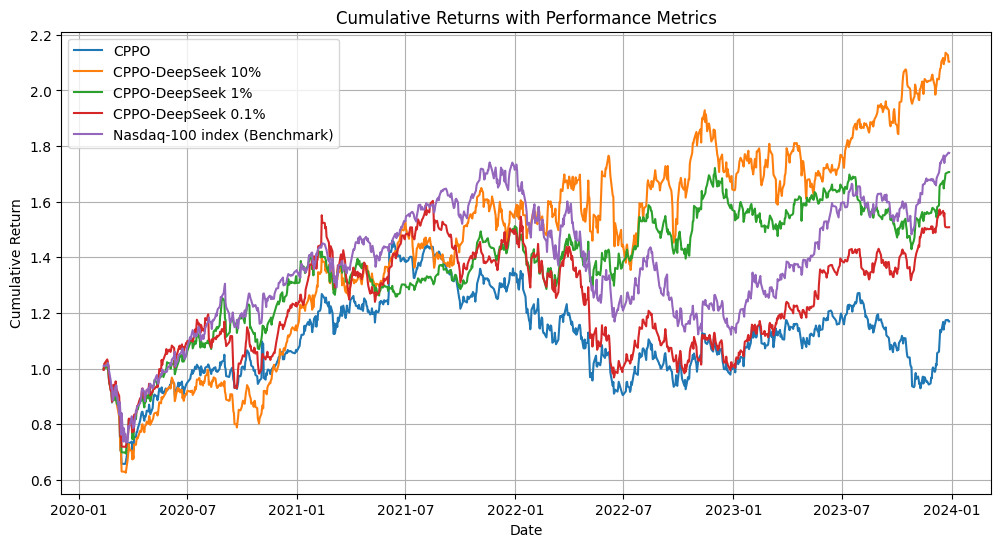}
    \caption{Impact of LLM infusion on CPPO}
    \label{fig8}
\end{figure}

\subsubsection{Figure \ref{fig8}:}

\begin{table}[h]
    \centering
    \resizebox{\columnwidth}{!}{%
    \begin{tabular}{lccc}
        \hline
        Model & Information Ratio & CVaR & Rachev Ratio \\
        \hline
        CPPO & -0.0148 & -0.0439 & 1.0404 \\
        CPPO-DeepSeek 10\% & 0.0078 & -0.0437 & 0.9818 \\
        CPPO-DeepSeek 1\% & -0.0032 & -0.0365 & 0.9573 \\
        CPPO-DeepSeek 0.1\% & -0.0060 & -0.0441 & 0.9789 \\
        \hline
    \end{tabular}%
    }
    \caption{Performance metrics for CPPO and CPPO-DeepSeek variants.}
    \label{tab:cppo_metrics}
\end{table}

For CPPO-DeepSeek, stronger LLM infusion improves performance

\section{Conclusion}
We propose LLM-infused RL agents for algorithmic trading that integrates stock trading recommendations  and risk assessments from news. Future work includes the following directions:

\begin{itemize}
    \item \textbf{Optimizing RAM usage:} Longer training requires more RAM. For instance, 500k training steps used 16 GB (+Swap memory), whereas 2 million steps required 128 GB (+Swap memory). Optimizing memory efficiency is crucial for scalability.
    
    \item \textbf{Reducing timescale:} Since markets react quickly and abruptly to events, reducing the timescale of decision-making could improve trading performance.
    
    \item \textbf{Improving news signal quality:} Enhancing the quality of news signals from the FNSPID dataset is an important direction to ensure better market performance.
\end{itemize}

\bibliographystyle{plainnat}
\bibliography{References}

\end{document}